\documentclass{aa}

\usepackage[T1]{fontenc}
\usepackage{txfonts}
\usepackage{amsmath}
\usepackage{amsfonts}
\usepackage{xcolor}
\usepackage{url}
\usepackage{bm}
\usepackage{graphicx}
\usepackage{amssymb}
\usepackage{soul}
\usepackage{booktabs}
\usepackage{array}
\usepackage[utf8]{inputenc}
\inputencoding{latin1}
\inputencoding{utf8}
\usepackage{appendix}
\usepackage{placeins}
\usepackage{multicol}

\def\lsim{ \lower .75ex\hbox{$\sim$} \llap{\raise .27ex \hbox{$<$}} }
\def\gsim{ \lower .75ex \hbox{$\sim$} \llap{\raise .27ex \hbox{$>$}} }

\newcommand{\fig}[1]{Fig.~\ref{fig:#1}}
\newcommand{\eq}[1]{Eq.~(\ref{eq:#1})}
\newcommand{\unit}[1]{\nobreak{\mathrm{\;#1}}} 

\usepackage[colorlinks=true,linkcolor=blue,citecolor=blue]{hyperref}

\title{Ultra-extreme high-frequency-peaked BL Lacs: A potential population of MeV synchrotron blazars} 

\author{
A. Sciaccaluga\inst{1}
\and 
F. Tavecchio\inst{1}
\and
T. Sbarrato\inst{1}
\and 
G. Bonnoli\inst{1}
}

\institute{
INAF -- Osservatorio Astronomico di Brera, Via E. Bianchi 46, I-23807 Merate, Italy\\
\email{alberto.sciaccaluga@inaf.it}
}
\date{}

\titlerunning{}

\begin{document}

\abstract{We investigate the potential existence of a new population of BL Lacs, called ultra-extreme high-energy-peaked BL Lacs (UEHBLs), whose synchrotron emission component peaks in the MeV band, extending the blazar sequence beyond its current limit. To model the spectral energy distribution of these new sources, we applied the hybrid shock-turbulence acceleration framework previously developed for extreme high-frequency-peaked BL Lacs. We present three representative realizations that produce synchrotron peaks between $0.2$ and $2$ MeV and evaluate their multiwavelength signatures. Our results show that UEHBLs would be undetectable with current GeV (Fermi) and future TeV (CTA) facilities due to severe Klein-Nishina suppression of inverse-Compton scattering, but are ideal targets for proposed MeV missions such as COSI, AMEGO-X, and e-ASTROGAM. We further identified a sample of hard X-ray sources from the Swift-BAT catalogs that exhibit the spectral properties expected of UEHBLs, representing promising follow-up targets of this population. If confirmed, UEHBLs would provide unique insight into particle acceleration in relativistic jets, imposing strong constraints on the maximum achievable electron energies. We also discuss the expected polarization and variability signatures, including the possibility of synchrotron-driven thermal instabilities leading to MeV flares. These findings underscore the critical importance of the MeV band for discovering and characterizing the most extreme accelerators among blazars.}


\maketitle

\section{Introduction}
\label{section:introduction}

Relativistic jets launched by active galactic nuclei (AGNs) emit radiation across the entire electromagnetic spectrum, from radio waves to gamma rays, and possibly also produce neutrinos. These jets provide unique laboratories for studying black hole physics, particle acceleration, and high-energy emission processes \citep[e.g.,][]{blandford+19}.  
Blazars are among the best targets for investigating jet physics. They are radio-loud AGNs whose relativistic jet is oriented close to our line of sight. Because of this alignment, the nonthermal jet emission is strongly amplified by relativistic beaming and dominates the spectral energy distribution (SED) \citep[e.g.,][]{romero+17, bottcher19}.

The SED of a blazar typically exhibits two broad humps. The low-energy hump peaks between the infrared and X-ray bands and is generally attributed to synchrotron emission from relativistic leptons. The high-energy hump, which peaks from MeV to TeV energies, is less well understood, and its origin remains debated. In the leptonic scenario, it is explained as the result of inverse-Compton scattering, where relativistic leptons upscatter photons from either their own synchrotron radiation, also called synchrotron self-Compton (SSC), or from external radiation fields \citep[e.g.,][]{maraschi+92, sikora+94, ghisellini+98}. Alternatively, hadronic models attribute the high-energy component to processes involving relativistic protons, such as direct proton synchrotron emission or synchrotron radiation from secondary particles produced in hadronic interactions \citep[e.g.,][]{cerruti20, sol&zech22}. The hadronic picture is supported by the possible association of some blazars with high-energy neutrino detections \citep[e.g.,][]{icecube+18}.

The positions of the two SED peaks are used to classify blazars, giving rise to the so-called blazar sequence, in which the peak frequencies are anticorrelated with the bolometric luminosity \citep{fossati+98, donato+01, ghisellini+17}. At the low-luminosity end of this sequence lie extreme high-frequency-peaked BL Lacs (EHBLs), which are the most efficient particle accelerators among blazars \citep[for a review,][]{biteau+20}. 

In the seminal paper by \cite{ghisellini99}, the author proposed extending the blazar sequence beyond its previously assumed limits, suggesting the existence of a new BL Lac population whose synchrotron spectrum peaks in the MeV band. In principle, there is no strict theoretical bound on the maximum synchrotron photon energy, except for the so-called synchrotron burn-off limit. If nonthermal particles are accelerated at shocks (or, more generally, by a gyroresonant process), the minimum acceleration timescale corresponds to the gyration time, $t_g = 2\pi r_g/c$, where $r_g$ is the particle gyroradius. This process is constrained by synchrotron cooling, which imposes an upper limit on the achievable photon energy that is independent of the magnetic field strength. For a relativistically beamed source, the maximum synchrotron photon energy can reach $\sim 25\,\delta \unit{MeV}$, comfortably exceeding the energies observed in EHBLs \citep[e.g.,][]{guilbert+83, dejager+96}. Specifically for high-energy BL Lacs, assuming that the acceleration timescale is a multiple of the gyration time, $t_\mathrm{acc} = \eta t_g$, leads to very high values, $\eta \sim 10^5$, implying that shocks are rather slow accelerators \citep[e.g.,][]{inoue&takahara96, garson+10}. 

A natural starting point for modeling MeV-peaking BL Lacs is one of the competing frameworks proposed for EHBLs. Owing to their peculiar spectral features, namely the large separation between the two peaks and their steep gamma-ray spectrum, the standard one-zone leptonic model typically requires unusually low magnetic fields ($B < 10 \unit{mG}$) and very high average electron Lorentz factors ($\bar{\gamma} \sim 10^3 - 10^4$). These values place EHBLs as clear outliers with respect to the parameter ranges inferred for other BL Lacs \citep{tavecchio+10, kaufmann+11, costamante+18}. To account for these differences, several alternative scenarios have been proposed, including a Maxwellian-like electron distribution \citep{lefa+11}, a beam of high-energy hadrons \citep{essey&kusenko10}, emission from a large-scale jet \citep{aharonian+08}, lepto-hadronic models \citep{cerruti+15}, and multiple shock acceleration \citep{zech&lemoine21}. 

In \cite{sciaccaluga&tavecchio22} and \cite{sciaccaluga+24}, we presented a novel framework for EHBLs based on the combined action of shock and turbulence acceleration. In this scenario, nonthermal leptons are initially accelerated by a shock and are subsequently energized by the downstream turbulence. 

All EHBL models can, in principle, be extended to even higher energies, potentially giving rise to a new population of BL Lacs. If these extreme sources are indeed observed, they would provide a unique laboratory for testing competing acceleration and emission scenarios. 

BL Lacs whose synchrotron hump peaks in the MeV band, here termed ultra-extreme high-frequency-peaked BL Lacs (UEHBLs), would be ideal targets for the upcoming missions designed to explore this energy range (for a review about extragalactic sources, see \citealt{sbarrato+25}). Historically, the $0.1 \unit{MeV}$–$100 \unit{MeV}$ band has been poorly sampled: only a few missions have partially covered it and did so with limited sensitivity. For this reason, this portion of the electromagnetic spectrum is commonly referred to as the MeV gap. In recent years, several satellites have been proposed to bridge this gap, spanning small-, medium-, and large-scale missions, such as the Compton Spectrometer and Imager \citep[COSI,][]{tomsick+22},  All-sky Medium Energy Gamma-ray Observatory eXplorer \citep[AMEGO-X,][]{caputo+22}, and enhanced ASTROGAM \citep[e-ASTROGAM,][]{deangelis+18}.

If UEHBLs do exist, they may have already left observable signatures in neighboring energy bands. For instance, in the hard X-rays, satellites such as Swift Burst Alert Telescope \citep[BAT,][]{barthelmy+05} provide excellent sky coverage and sensitivity. Interestingly, the BAT catalogs include a number of sources without clear counterparts, which might represent promising candidates for this hypothesized BL Lac population \citep{oh+18, lien+25}.

The paper is organized as follows. In Section~\ref{section:model} we describe the hybrid shock–turbulence acceleration model. Section~\ref{section:results} reports and discusses three representative realizations of the model. In Section~\ref{section:discussion} we examine the potential temporal and polarimetric characteristics of UEHBLs and discuss the implications for possible acceleration and emission mechanisms. Throughout the paper, the following cosmological parameters are assumed: $H_0 = 70 \unit{km s^{-1} Mpc^{-1}}$, $\Omega_M = 0.3, and \Omega_\Lambda = 0.7$.

\section{Shock-turbulence model}
\label{section:model}

In this section, we briefly describe our shock-turbulence acceleration model. This is one possible EHBL model that might be extended to UEHBLs (for more details, see \citealt{sciaccaluga&tavecchio22} and \citealt{sciaccaluga+24}). As mentioned in the introduction, \cite{zech&lemoine21} proposed a model based on multiple shock acceleration. Their idea relied on the fact that recollimation (or, more generally, standing) shocks occur in series, as demonstrated by several 2D fluid simulations \citep[e.g.,][]{gomez+95, mizuno+15}. However, 3D fluid simulations revealed that the flow is subject to instabilities that evolve into turbulence and ultimately disrupt the cycle of recollimation and reflection shocks \citep[e.g.,][]{matsumoto&masada13, gourgouliatos&komissarov18, boula+25, hu+25, costa+25}. For this reason, we assumed that particles are first accelerated by a shock and then further energized by the downstream turbulence. 

We assumed that the emitting region is the turbulent downstream of a recollimation shock, and we adopted a leaky-box spatially averaged approach, so that our scenario effectively was a one-zone leptonic model. The temporal evolution of the electrons and turbulence is governed by two coupled Fokker–Planck equations \citep[e.g.,][]{eilek79, miller+96, kakuwa16, gong+25},
\begin{equation}
\left \{
\begin{aligned}
    & \frac{\partial f}{\partial t} = \frac{1}{p^2} \frac{\partial}{\partial p} \left [ p^2 D_p \frac{\partial f}{\partial p} + p^2 \left ( \frac{\partial p}{\partial t} \right )_{\text{rad}} f \right ] - \frac{f}{t_\text{esc}} +I_f \\
    & \frac{\partial W}{\partial t} = \frac{\partial}{\partial k} \left [ k^2 D_k \frac{\partial}{\partial k} \left(\frac{W}{k^2} \right) \right ] - \frac{W}{t_\text{dam}} + I_w   
\end{aligned}
\label{eq:system}
\right ., 
\end{equation}
where $p$ is the electron momentum, $k$ the fluctuation wavenumber, $f(p,t)$ the electron isotropic phase-space density, and $W(k,t)$ the turbulence energy density per unit of wavenumber. \eq{system} describes the interplay between particle processes (resonant acceleration, cooling, escape, and injection) and turbulence processes (cascading, damping, and injection). All quantities are defined in the comoving frame of the emission region. The expressions for particle acceleration and turbulence damping are derived from quasi-linear theory, although alternative approaches exist \citep[e.g.,][]{lemoine+24}.

The diffusion coefficient of electrons is given by
\begin{equation}
    D_p = \frac{p^2 \,\beta_a^2 \, c}{U_B \, r_g^2} \int_{k_\text{res}} \frac{W_B}{k}\,dk,
    \label{eq:Dp}
\end{equation}
where $\beta_a$ is the dimensionless Alfvén speed, $U_B = B^2/8\pi$ is the magnetic energy density, $r_g$ is the electron gyroradius, $W_B \approx W/2$ is the magnetic component of the turbulence energy spectrum, and $k_\text{res} = 1/r_g$ is the resonant wavenumber.

The electron cooling term accounts for synchrotron and inverse-Compton contributions, following standard formulae from the literature \citep[e.g.,][]{jones65, chiaberge&ghisellini99}.

The electron escape time is equal to 
\begin{equation}
    t_\text{esc} = \frac{R}{c} + \frac{R^2}{\kappa_\parallel},
\end{equation}
where $R$ is the emission region radius, and $\kappa_\parallel = c r_g /9 \zeta(k_\text{res})$ is the spatial diffusion coefficient along the magnetic field, with $\zeta(k) = kW_B/U_B$ the relative amplitude of the turbulent magnetic field energy density for a given $k$. When the mean free path is large, the escape time is essentially the geometric escape time. Conversely, when the mean free path is small, turbulence traps particles in the emission region, leading to a longer escape time.

Before gaining energy through turbulence, particles are initially accelerated at the shock, which thus acts as the injector for the emission region. The electron injection number density per unit of Lorentz factor $I_{n}$ is given by
\begin{equation}
    I_n = I_{n,0} \,\gamma^{-p}\, e^{-\frac{\gamma}{\gamma_\text{cut}}} \quad \text{with}\ \, \gamma > \gamma_\mathrm{min},
\end{equation}
where $\gamma$ denotes the electron Lorentz factor, $I_{n,0}$ is the injection normalization, $p=2$ is the power-law slope, and $\gamma_\mathrm{min} = 10^3$ and $\gamma_\mathrm{cut} = 10^5$ are the minimum and cutoff Lorentz factor, respectively, of the injected electron distribution. We note that $I_n = 4\pi p^2 m_e c \, I_f$, where $I_f$ is the corresponding injection distribution in phase space. The slope and cutoff of the injection were determined from recent simulations of diffusive shock acceleration for weakly magnetized shocks \citep[e.g.,][]{vanthieghem+20, zech&lemoine21}. The injection was normalized to the injected electron power $P_n$,
\begin{equation}
     P_n = V \int \gamma \, m_e\, c^2\, I_n \,d\gamma,
\end{equation}
where $V = 10\pi R^3$ is the emission region volume, modeled as a cylinder with radius $R$ and length $10\,R$. The length of the emission volume, related to the region where the instability develops and triggers turbulence in the plasma, was roughly estimated on fluid simulations \citep[e.g.,][]{matsumoto+21, boula+25, costa+25}. We expect that within this distance, the magnetic field decay and the adiabatic losses effectively quench the emission. 

In a Kolmogorov phenomenology, the diffusion coefficient of turbulence is given by
\begin{equation}
    D_k = k^3\, \beta_a\, c \sqrt{\frac{k W}{2 U_B}}.
\end{equation}
Without strong damping, and given this diffusion coefficient with continuous injection, $W(k)$ would evolve toward the standard Kolmogorov spectrum, $W(k) \propto k^{-5/3}$ \citep{zhou&matthaeus90}.

The turbulence damping time is equal to 
\begin{equation}
   t_\text{dam} = -\frac{4\pi e^2 \beta_a^2 }{m_e c k} \int_{\gamma_\text{res}} \gamma^2 \frac{\partial}{\partial \gamma} \left ( \frac{n_e}{\gamma^2} \right ) d\gamma,
\end{equation}
where $n_e(\gamma) = 4\pi m_e c p^2 f(p)$ is the electron number density per unit of Lorentz factor, and $\gamma_\text{res}$ is the resonant Lorentz factor, defined as $k = 1/r_g(\gamma_\text{res})$. Damping is determined by enforcing energy conservation, so that the energy driving electron acceleration is removed from the turbulence. 

The turbulence injection term is equal to
\begin{equation}
    I_W = I_{W,0}\, \delta (k-k_0),
\end{equation}
where $\delta$ is the Dirac function, $I_{w,0}$ is the normalization, and $k_0 = 1/L$ is the injection wavenumber, with $L = R/10$. The injection is normalized to the injected turbulence power $P_w$,
\begin{equation}
     P_w = V \int I_W \,dk.
\end{equation}

The radiative output of the emission region was computed using the SSC model, following the standard formulae in the literature \citep[e.g.,][]{jones68, blumenthal&gould70, ghisellini+88}. For the treatment of relativistic beaming, we adopted the usual blob amplification formula \citep[see Appendix C of][]{zech&lemoine21}.

We employed the Chang–Cooper algorithm \citep{chang&cooper70} on a logarithmic grid, using $20$ points per decade for momentum, wavenumber, and frequency. The system was evolved over a time interval of $10\,R/c$ with $100$ time steps.

\section{Results}
\label{section:results}

\begin{figure*}[htbp]
\centering
\includegraphics[width=0.75\textwidth]{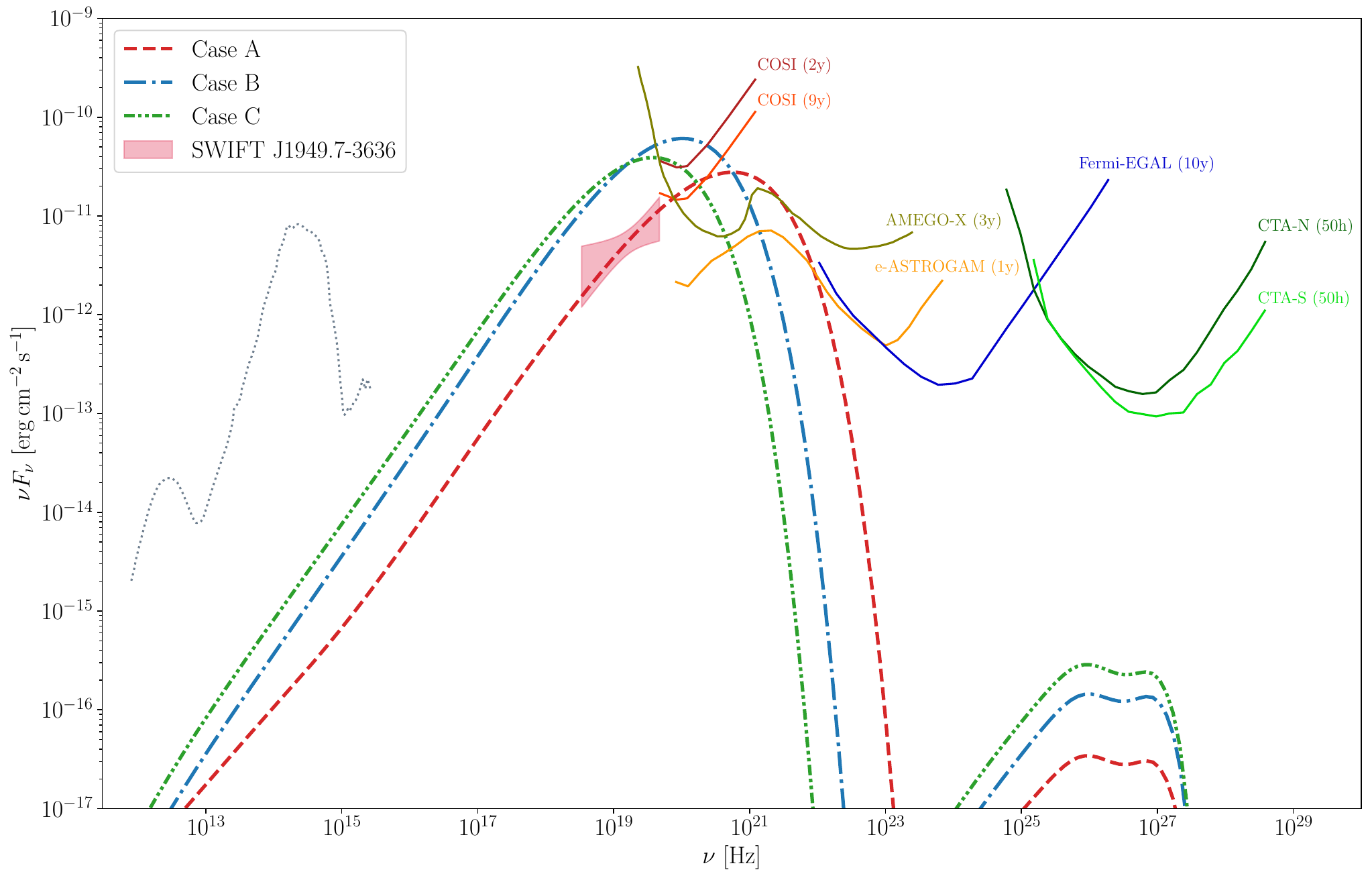}
\caption{Spectral energy densities for three model realizations: Case A (dashed red), case B (dash–dotted blue), and case C (dash–double–dotted green), including attenuation from extragalactic background light absorption \citep{franceschini&rodighero17}. The solid dark and light red lines show COSI 2- and 9-year sensitivities \citep{tomsick+22}, and the dark yellow and orange lines show AMEGO-X (3 yr; \citealt{caputo+22}) and e-ASTROGAM (1 yr; \citealt{deangelis+18}) sensitivities. The Fermi-LAT 10-year extragalactic sensitivity (taken from \url{https://www.slac.stanford.edu/exp/glast/groups/canda/lat_Performance.htm}) is shown as a solid blue line, and the CTA North and South 50-hour sensitivities (taken from \url{https://www.ctao.org/for-scientists/performance/}) are shown as dark and light green lines. The red shaded region marks the BAT-band flux of Swift J1949.7–3636 \citep{lien+25}, a candidate MeV-peaking BL Lac. For reference, an SED template of a giant elliptical galaxy taken from \citep{silva+98} is displayed in dotted grey, renormalized to the magnitude of the host galaxy of 1ES 0229+200 \citep{costamante+18}.}
\label{fig:seds_and_sensitivities}
\end{figure*}

In this section, we discuss some examples of realizations of the hybrid shock–turbulence model that produce BL Lacs with a spectral peak in the MeV band. In \fig{seds_and_sensitivities}, we present three representative realizations. The model depends on six parameters: the emission region radius $R$, the dimensionless Alfvén speed $\beta_a$, the magnetic field strength $B$, the injected electron power $P_e$, the injected turbulence power $P_w$, and the relativistic Doppler factor $\mathcal{D} = [\Gamma (1-\beta \cos{\theta_v})]^{-1}$ (where $\Gamma$ and $\beta$ denote the bulk Lorentz factor and the dimensionless velocity of the fluid, respectively, while $\theta_v$ is the observer’s viewing angle). The parameter sets corresponding to the three realizations are summarized in Tab.~\ref{tab:parameters}. In all cases, the redshift was fixed at $z = 0.14$, matching that of the prototypical EHBL 1ES 0229+200 \citep{woo05}. In addition to the input parameters, the table also lists two derived quantities we used to check the model consistency: the magnetization, $\sigma = \beta_a^2 / (1 - \beta_a^2)$, and the relative amplitude of turbulent magnetic fluctuations, $\delta B / B = \left(\int W_B \, dk / U_B\right)^{1/2}$.

\begin{table}[htbp]
\caption{Physical parameters and derived quantities for the three modeled cases}
\resizebox{\columnwidth}{!}{
\begin{tabular}{cccc}
\hline
\hline
Parameter                             & Case A               & Case B               & Case C               \\ \hline \addlinespace[2pt]
Emission region radius {[}cm{]}       & $1.0 \times 10^{16}$ & $1.0 \times 10^{16}$ & $1.0 \times 10^{16}$ \\
Magnetic field {[}G{]}                & $1.7 \times 10^{-2}$ & $3.0 \times 10^{-2}$ & $3.0 \times 10^{-2}$ \\
Dimensionless Alfvén speed            & $2.0 \times 10^{-1}$ & $2.0 \times 10^{-1}$ & $2.0 \times 10^{-1}$ \\
Injected electron power {[}erg/s{]}   & $1.0 \times 10^{37}$ & $3.0 \times 10^{36}$ & $3.0 \times 10^{36}$ \\
Injected turbulence power {[}erg/s{]} & $1.0 \times 10^{40}$ & $1.0 \times 10^{40}$ & $3.0 \times 10^{39}$ \\
Relativistic Doppler factor           & $1.6 \times 10^{1}$  & $2.0 \times 10^{1}$  & $2.5 \times 10^{1}$  \\ \hline \addlinespace[2pt]  
Magnetization                         & $4.2 \times 10^{-2}$ & $4.2 \times 10^{-2}$ & $4.2 \times 10^{-2}$ \\
Turbulence relative magnitude          & $2.0 \times 10^{-1}$ & $1.4 \times 10^{-1}$ & $9.4 \times 10^{-2}$ \\ \hline \hline
\end{tabular}
}
\label{tab:parameters}
\end{table}

In all three cases, the emission region radius was fixed at subparsec scales, as expected for BL Lacs \citep{tavecchio+98, tavecchio+10}. The dimensionless Alfvén speed was set to ensure low magnetization, which makes shock acceleration efficient and favors the development of instabilities and turbulence in the downstream region \citep[e.g.,][]{vanthieghem+20, matsumoto+21}. The remaining parameters were varied as needed. However, since we adopted the quasi-linear approximation, we verified that the relative turbulence amplitude remained small, tha tis, $\delta B / B \ll 1$. The time evolution of the electron and turbulence spectra for case A is reported in Appendix \ref{section:electron_turbulence_spectrum}. Cases B and C exhibit a similar behavior and are therefore not shown.

Case A provides the most efficient acceleration, with the low-energy bump peaking at $\sim 2 \unit{MeV}$. In the MeV band, the low-energy peak is fully detectable by AMEGO-X and e-ASTROGAM, with the latter also able to detect case A beyond the peak. COSI cannot observe the source within its nominal mission lifetime, namely 2 years, although case A would become detectable if the mission were extended. Detection by COSI would also be possible in the event of a strong flare. Although EHBLs, the closest known class of sources to those hypothesized here, do not exhibit strong flaring at the frequencies corresponding to the low-energy bump, radiative thermal instabilities could trigger such flares, as discussed in Section~\ref{section:discussion}.

The relative turbulence amplitude in case A is close to the nominal boundary of the quasilinear regime. However, it is possible to obtain spectra with the same synchrotron peak energy for lower levels of magnetic turbulence, that is, lower values of $\delta B / B$. As discussed in \cite{sciaccaluga+24}, the model parameter space is significantly degenerate, which allows different combinations of physical quantities to yield similar spectral outcomes. For example, we can increase the Alfvén speed parameter $\beta_a$ while reducing the injected turbulence power (see case D in Appendix~\ref{section:additional_cases}). This leads to a higher magnetization, with corresponding values as high as $\sigma \sim 0.1$, and a reduced turbulence level $\delta B / B \sim 0.1$. However, this relatively high magnetization might suppress the efficiency of shock acceleration and limit the development of downstream turbulence. Alternatively, a decrease in the injected turbulence power can be compensated for by an increase in the Doppler factor. This allows for configurations with $\delta B / B \sim 0.1$ that still produce a spectrum peaking in the same region as case A (see case E in Appendix~\ref{section:additional_cases}). These examples illustrate that the appearance of spectra is not uniquely tied to a narrow region of parameter space, but instead arises from a broader set of physically plausible configurations. A comprehensive and systematic exploration of the parameter degeneracy is beyond the scope of the present work and will be addressed in future studies.

Potential constraints might arise from the hard X-ray band. As mentioned in the introduction, valuable information is expected from Swift-BAT, which operates in the hard X-rays and offers excellent sky coverage. Recently published BAT catalogs \citep{oh+18, lien+25} contain several candidates that may represent UEHBLs. To identify such candidates, we filtered the BAT catalogs with the following criteria: i) integrated flux in the BAT band above $10^{-11} \unit{erg\, cm^{-2}\, s^{-1}}$, ii) photon index $\Gamma < 1.5$, iii) likely extragalactic origin (defined by a Galactic latitude $|b| > 10^\circ$), iv) absence of a clear counterpart in the soft X-ray and optical bands. The requirements on the flux and index in the hard X-rays were imposed to identify sources that might be detectable by the next generation of MeV telescopes. Applying these requirements, we identified ten possible sources according to the catalog classifications. Their number might drop to six when possible X-ray counterpart candidates are updated (see Appendix \ref{section:bat_candidates} for further details). In \fig{seds_and_sensitivities},
we plot the integrated BAT flux of one of the candidates, SWIFT J1949.7-3636. All BAT candidates lack a soft X-ray counterpart that would allow us to clearly associate them with sources in optical catalogs. This is consistent with case A, in which the soft X-ray flux is close to the sensitivity limit of current soft X-ray satellites. A potential approach to identifying a soft X-ray counterpart would be to perform a deep observation of the BAT error region using a soft X-ray telescope with a sufficiently large field of view, such as Swift X-Ray Telescope \citep[XRT,][]{burrows+05} and XMM-Newton \citep{jansen+01}. In the event of multiple detections, spectral characteristics (e.g., a hard photon index) might help us to distinguish among the possible candidates. These candidates could be FSRQs with their high-energy peak located in the MeV band. In this scenario, BAT would be detecting the rising part of the high-energy component. However, UEHBLs and MeV FSRQs are easily distinguishable. First, given a soft X-ray detection providing better localization, the optical counterpart should present broad emission lines in the case of FSRQs. Second, MeV FSRQs after the peak typically exhibit a power-law tail extending into the GeV band, which should be detectable by Fermi Large Area Telescope (LAT).

The case A parameters were selected to ensure consistency with the BAT data, but this requirement is highly restrictive. For realizations B and C, we therefore chose not to impose this constraint. From a theoretical perspective, there is no strict limitation on the properties of potential MeV BL Lacs, and intermediate cases between EHBLs and case A should, in principle, also be possible. Moreover, as previously highlighted, these sources may not remain stable over time. Case A is consistent with a BAT source that has been monitored for 157 months. However, during this period, the source may have experienced flaring episodes, during which its brightness exceeded the average flux measured by BAT. 

Case B has a low-energy bump peaking at $\sim 0.5 \unit{MeV}$, and it is the brightest candidate in the MeV band, potentially detectable by COSI within its nominal mission lifetime. Case C, with a peak at $0.2 \unit{MeV}$, represents the scenario most closely resembling EHBLs. While COSI could still detect case C within its lifetime, the signal would lie near the edge of its sensitivity range, allowing observation only just beyond the peak. The superior sensitivities of AMEGO-X and e-ASTROGAM would enable them to probe the peak and higher energies in cases B and C. 

None of the three cases is observable by Fermi or the Cherenkov Telescope Array (CTA) because the scattering cross section in the Klein–Nishina regime is severely suppressed in a full leptonic scenario (see Section~\ref{section:discussion} for further details on a possible hadronic component).

Finally, all three cases, the optical band is dominated by the galaxy emission, even more than in EHBLs, as shown in \fig{seds_and_sensitivities}. A more accurate localization, for example, using soft X-ray observations, would enable us to detect the host galaxy of these sources. 

\section{Discussion}
\label{section:discussion}

We demonstrated that the hybrid shock–turbulence scenario, one of the possible models for EHBLs, can also account for UEHBLs. We presented three representative model realizations and discussed their detectability with proposed MeV observatories and with existing telescopes at other wavelengths. In addition, we identified ten potential candidates from the BAT catalogs.

Proposed MeV satellites such as COSI, AMEGO-X, and e-ASTROGAM are essential for detecting these new sources. Their observations would provide valuable insights into particle acceleration. Previous studies have already shown that MeV observations of powerful FSRQ (sources at the opposite extreme of the blazar sequence with respect to EHBLs) could reveal the presence of a thermal component in the particle spectrum in addition to the standard nonthermal power law. This would indicate that shocks are the underlying acceleration mechanism \citep{tavecchio+25}. On the other hand, the detection of BL Lacs peaking in the MeV band would imply that electrons can achieve extremely high Lorentz factors. This condition might help us to  further distinguish among the different particle acceleration mechanisms proposed in the literature. Conversely, a non-detection of such sources could be interpreted in two ways. First, particle acceleration, regardless of the underlying mechanism, must proceed relatively slowly, as discussed in Section~\ref{section:introduction}. Second, UEHBLs may simply be too faint to be detected by the proposed MeV missions.

In addition to the spectral properties of UEHBLs, we are also interested in their polarimetric and temporal features. In the MeV band, we expect relatively high polarization degrees ($\gtrsim 40\%$), following the trend observed in HBLs and EHBLs \citep[e.g.,][]{digesu+22, liodakis+22, ehlert+23, kouch+24}, where the polarization is strongly chromatic, that is, the degree of polarization increases with frequency. Based on the flux predicted by our model (excluding strong flaring states), the polarization of UEHBLs would remain undetectable for COSI and AMEGO-X, while e-ASTROGAM might measure it, depending on the actual polarization degree. 

It is interesting to note that UEHBLs might represent an exceptional class of sources. In contrast to standard blazars, their emission would be dominated by synchrotron radiation, since inverse-Compton contribution is strongly quenched by Klein-Nishina suppression. This peculiarity might affect the variability properties of these sources. A source dominated by the pressure of nonthermal electrons and whose cooling is dominated by synchrotron losses might indeed be subject to the radiative thermal instability discussed by \cite{marscher80}. A small increase in the magnetic field in limited portions of the flow can, in fact, lead to a collapse of the region due to the increased cooling rate of the relativistic electrons and consequent loss in pressure. The rapid radiative losses suffered by the electrons during the compression would produce a rapid increase in the emissivity, with flares potentially detectable in the MeV band.

As already outlined, in a purely leptonic scenario, the inverse-Compton emission is expected to be suppressed, since most electrons scatter photons in the Klein–Nishina regime. Consequently, as illustrated in \fig{seds_and_sensitivities}, UEHBLs are expected to remain undetectable by CTA. However, we cannot exclude that other mechanisms can result in a detectable flux even at these energies. First, the emission might originate from a nonthermal hadronic component producing synchrotron radiation at TeV energies \citep[e.g.,][]{mannheim93, aharonian00, mucke+01}. Alternatively, a second emission region farther downstream in the jet might exist, where nonthermal electrons generate infrared/radio photons. These photons might then be upscattered by the nonthermal electrons in the primary emission region in the Thomson regime, giving rise to a bright inverse-Compton bump detectable by CTA. Variability studies might help us to distinguish between these two scenarios. A rapid variability on timescales of hours would disfavor the hadronic model, since the synchrotron cooling timescale for protons is expected to be orders of magnitude longer.

\begin{acknowledgements}
This work has been funded by ASI under contract 2024-11-HH.0. We acknowledge financial support from an INAF Theory Grant 2024 (PI F. Tavecchio) and the European Union-Next Generation EU, PRIN 2022 RFF M4C21.1 (2022C9TNNX).
\end{acknowledgements}

\bibliographystyle{aa}
\bibliography{bibliography}

\appendix

\section{Electron and turbulence spectra}
\label{section:electron_turbulence_spectrum}

\begin{figure}[htbp]
\centering
\includegraphics[width=\columnwidth]{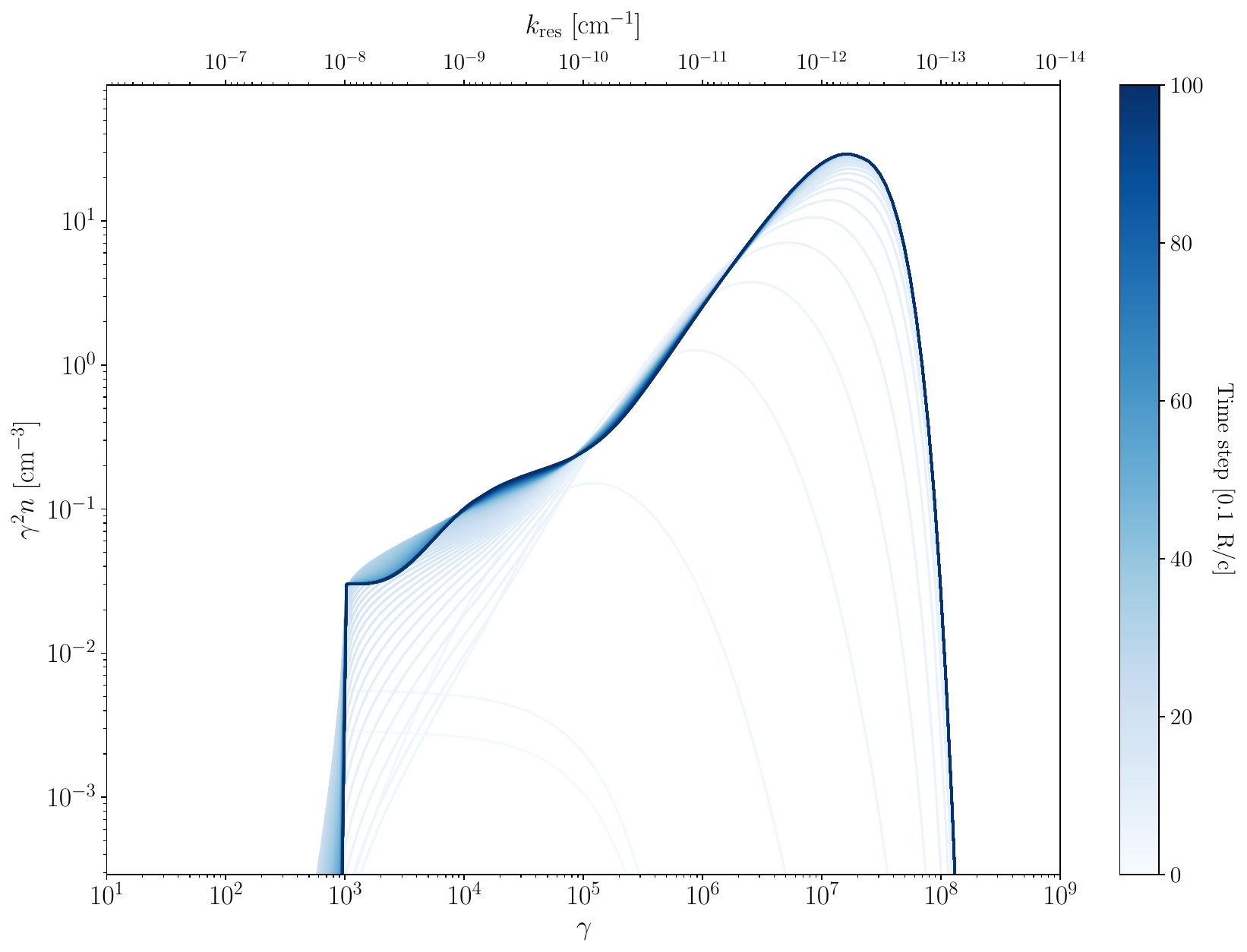}
\caption{Time evolution of the electron number density per unit of Lorentz factor as a function of Lorentz factor. The top axis reports the corresponding resonant wavenumber.}
\label{fig:electrons_A}
\end{figure}

\begin{figure}[htbp]
\centering
\includegraphics[width=\columnwidth]{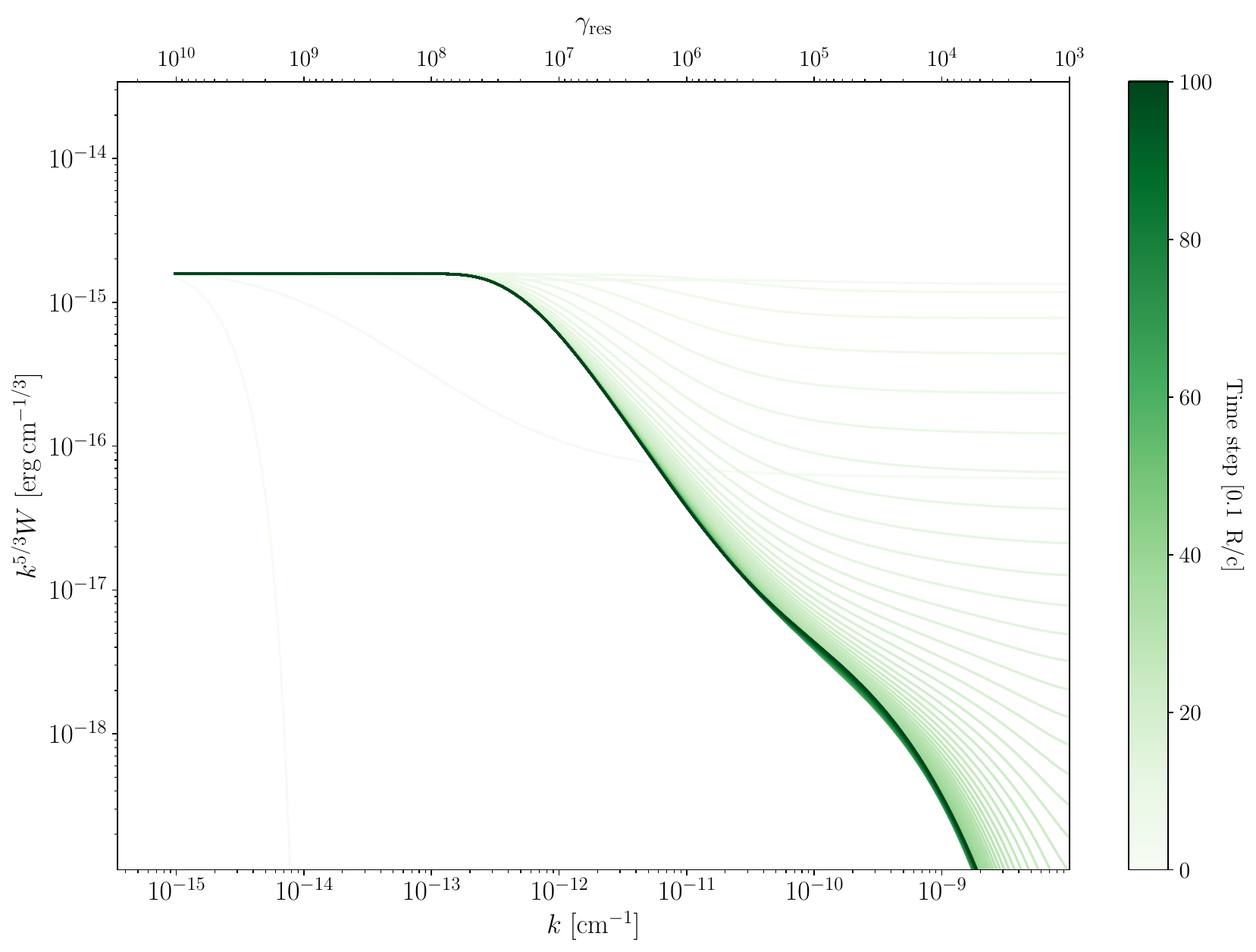}
\caption{Time evolution of the turbulence energy density per unit of wavenumber as a function of wavenumber. The top axis reports the corresponding resonant Lorentz factor.}
\label{fig:turbulence_A}
\end{figure}

Figures~\ref{fig:electrons_A} and \ref{fig:turbulence_A} show the time evolution of the electron and turbulence spectra of Case A, respectively. Electrons, injected into the downstream region with Lorentz factors in the range $10^3 < \gamma \lesssim 10^5$, are accelerated by the turbulence up to $\gamma \gtrsim 10^7$, where radiative cooling becomes dominant. During the acceleration, electrons extract energy from the turbulence, leading to strong damping at wavenumbers corresponding to the Lorentz factors where acceleration is most efficient. At lower wavenumbers, where the turbulence cascade dominates over electron damping, turbulence follows the standard Kolmogorov spectrum. The damping has further consequences. Initially, electrons tend to accumulate at high Lorentz factors. However, as the turbulence is progressively damped, the acceleration efficiency decreases. Consequently, at later times, the injected population becomes dominant again, flattening the spectrum at low Lorentz factors. 

\section{BAT candidates}
\label{section:bat_candidates}

\begin{table}[htbp]
\caption{Candidate BAT sources with their corresponding right ascension (RA), declination (Dec), observed flux, photon index, and catalogs association class.}
\resizebox{\columnwidth}{!}{
\begin{tabular}{cccccc}
\hline \hline \addlinespace[2pt]
BAT Name           & RA [$\mathrm{deg}$]      & Dec [$\mathrm{deg}$]      & Flux [$10^{-12} \unit{erg\, cm^{-2}\, s^{-1}}$]     & Photon index               &Catalogs association class                                                   \\ \hline \addlinespace[4pt]
SWIFT J0007.8-4133$^{*}$ & 1.9545  & -41.3559 & $10.61_{-3.52}^{+3.95}$                             & $1.41_{-0.51}^{+0.50}$     & {\small No archival soft X-ray image}                             \\ \addlinespace[4pt]
SWIFT J0656.0-6560$^{*}$ & 103.989 & -65.5981 & $10.02_{-3.18}^{+3.52}$                             & $1.42_{-0.49}^{+0.49}$     & {\small No archival soft X-ray image}                             \\ \addlinespace[4pt]
SWIFT J0722.5+2121 & 110.668 & 21.4408  & $12.03_{-4.49}^{+5.02}$                             & $1.21_{-0.63}^{+0.61}$     & {\small No soft X-ray detection}                                   \\ \addlinespace[4pt]
SWIFT J1334.1-3842 & 203.437 & -38.4501 & $14.17_{-4.19}^{+4.57}$                             & $1.35_{-0.48}^{+0.47}$     & {\small No archival soft X-ray image}                             \\ \addlinespace[4pt]
SWIFT J0045.9+3931 & 11.5092 & 39.531   & $11.95_{-3.58}^{+3.88}$                             & $1.27_{-0.46}^{+0.47}$     & {\small No likely soft X-ray counterpart}                          \\ \addlinespace[4pt]
SWIFT J0106.1+4818 & 16.5232 & 48.2949  & $11.59_{-3.40}^{+3.71}$                             & $1.32_{-0.43}^{+0.42}$     & {\small Possible Blazar}                                           \\ \addlinespace[4pt]
SWIFT J0243.2-0553$^{*}$ & 40.8043 & -5.8867  & $10.6_{-3.61}^{+4.12}$                              & $1.48_{-0.52}^{+0.53}$     & {\small Unavailable/unchecked soft X-ray image or crowded region}  \\ \addlinespace[4pt]
SWIFT J0449.3+6356 & 72.3282 & 63.9334  & $14.12_{-3.61}^{+3.93}$                             & $1.38_{-0.40}^{+0.39}$     & {\small No likely soft X-ray counterpart}                          \\ \addlinespace[4pt]
SWIFT J1026.3+4536 & 156.586 & 45.5992  & $10.75_{-3.23}^{+3.53}$                             & $1.26_{-0.47}^{+0.46}$     & {\small Unavailable/unchecked soft X-ray image or crowded region}  \\ \addlinespace[4pt]
SWIFT J1949.7-3636$^{*}$ & 297.436 & -36.5979 & $14.07_{-3.71}^{+4.04}$                             & $1.49_{-0.41}^{+0.41}$     & {\small Unavailable/unchecked soft X-ray image or crowded region}  \\ \addlinespace[4pt]
\hline \hline
\end{tabular}
}
\label{tab:bat_candidates}
\\[2mm] 
\parbox{\linewidth}{\small \textbf{Notes.} Further details on the counterpart association procedure followed by the catalogs are available in Section~2.1 of \citealt{oh+18} and Section~2 of \citealt{lien+25}. An asterisk ($^{*}$) marks sources for which a possible soft X-ray counterpart was identified upon further inspection of archival data described in the text.}
\end{table}

We further checked the existence of soft X-ray counterparts for these 10 sources using archival images and catalogs provided by Swift-XRT, Chandra, XMM-Newton, ROSAT, and eROSITA. We confirm the absence of any likely X-ray counterpart for all sources except 4, which show a non-negligible association probability with sources characterized by X-ray fluxes close to the limiting value defined by \citet{lien+25}. Specifically, SWIFT J0007.8-4133 lies 8 arcmin from MCG-07-01-011, a Seyfert 2 galaxy detected by XMM-Newton, Swift-XRT, and eROSITA; SWIFT J0656.0-6560 is 6 arcmin from Fairall 0265, a Seyfert 1 galaxy detected by ROSAT and Swift-XRT; SWIFT J0243.2-0553 is located 2.3 arcmin from the known $\gamma$-ray emitting blazar PKS 0240-060; and SWIFT J1949.7-3636 is close to a ROSAT-detected source with no identified optical counterpart.

\section{Additional cases}
\label{section:additional_cases}

\begin{figure}[htbp]
\centering
\includegraphics[width=\columnwidth]{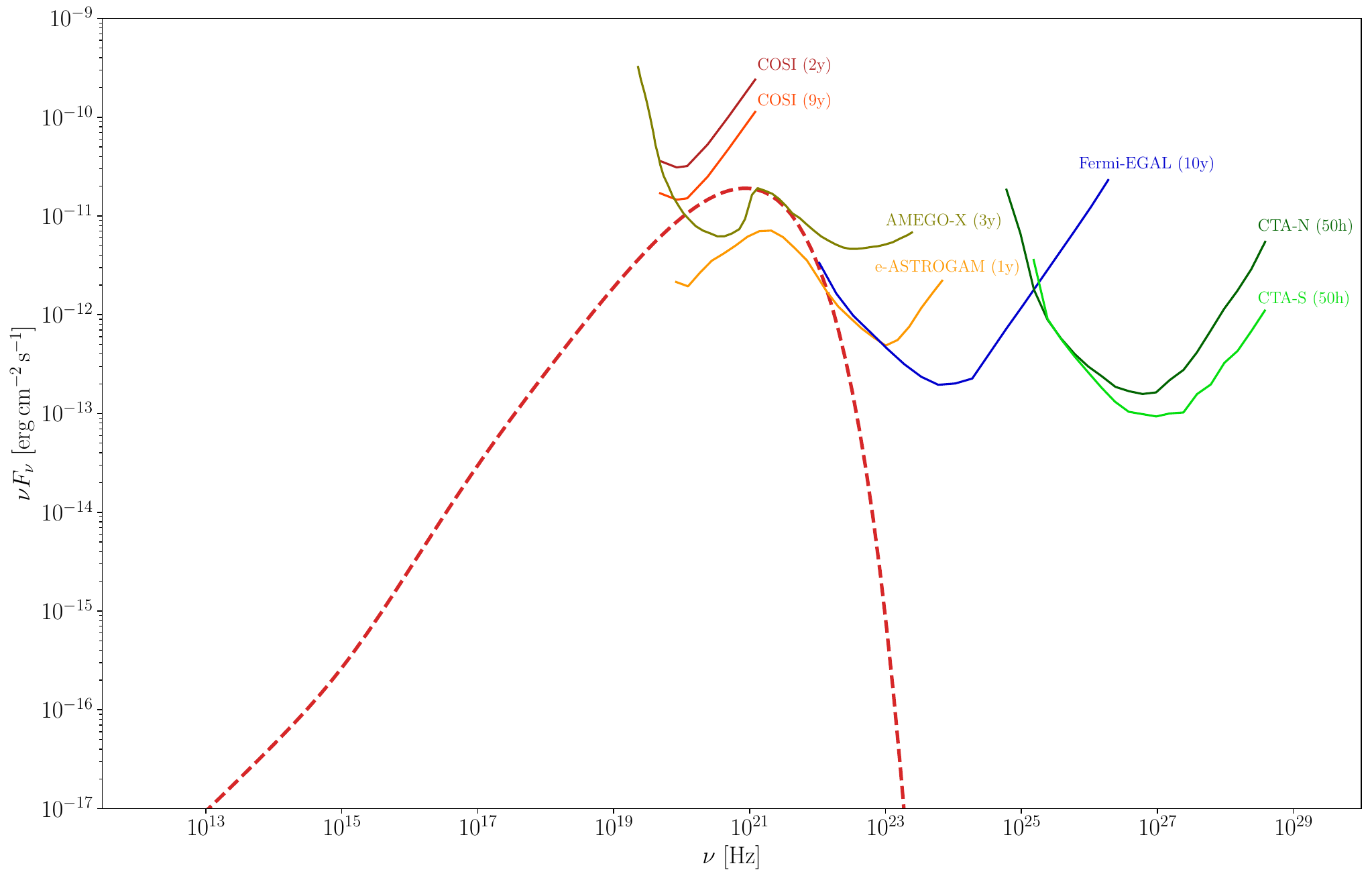}
\caption{The parameters of Case D are $R = 1.0 \times 10^{16}\,\mathrm{cm}$, $B = 1.7 \times 10^{-2}\,\mathrm{G}$, $\beta_a = 3.0 \times 10^{-1}$, $P_n = 1.0 \times 10^{37}\,\mathrm{erg/s}$, $P_w = 6.7 \times 10^{39}\,\mathrm{erg/s}$, $\mathcal{D} = 1.6 \times 10^{1}$. The magnetization is $\sigma = 1.0 \times 10^{-1}$, while the relative amplitude of turbulent magnetic fluctuations is $\delta B/B = 1.5 \times 10^{-1}$}
\end{figure}

\begin{figure}[htbp]
\centering
\includegraphics[width=\columnwidth]{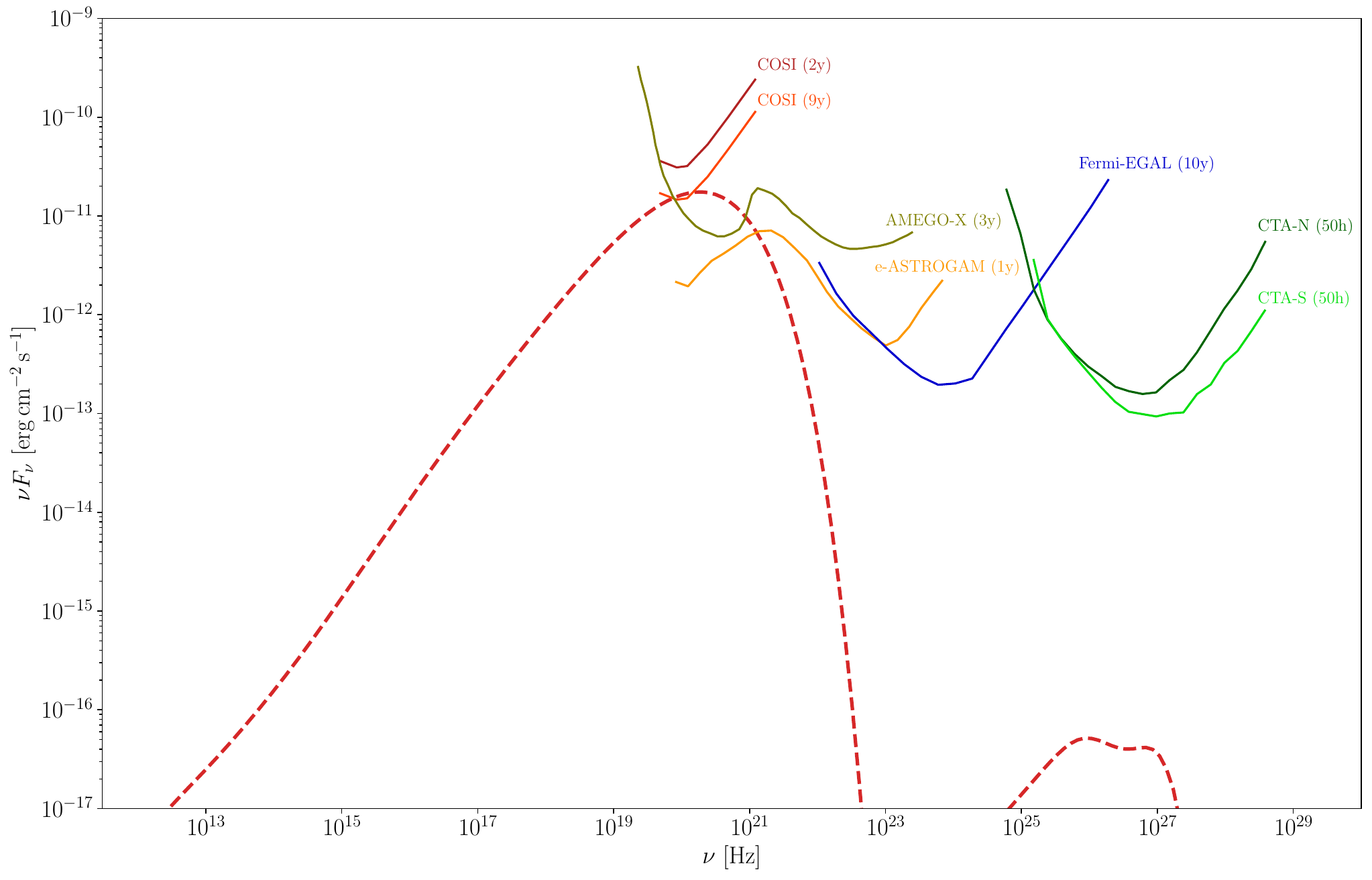}
\caption{The parameters of Case E are $R = 1.0 \times 10^{16}\,\mathrm{cm}$, $B = 1.7 \times 10^{-2}\,\mathrm{G}$, $\beta_a = 2.0 \times 10^{-1}$, $P_n = 1.0 \times 10^{37}\,\mathrm{erg/s}$, $P_w = 2.9 \times 10^{39}\,\mathrm{erg/s}$, $\mathcal{D} = 2.0 \times 10^{1}$. The magnetization is $\sigma = 4.2 \times 10^{-2}$, while the relative amplitude of turbulent magnetic fluctuations is $\delta B/B = 1.3 \times 10^{-1}$}
\end{figure}
\end{document}